# Two New Cataclysmic Variables in Lyra

D. V. Denisenko

[1] Space Research Institute, Moscow, Russia; e-mail: denis@hea.iki.rssi.ru

I report on the discovery of two cataclysmic variables in the same field in Lyra originally identified from their magnitudes in USNO-B1.0 catalogue and Palomar images. The historical light curves were analyzed on 300+ photographic plates from Moscow collection covering 35 years of observations. One of two stars, USNO-B1.0 1320-0390658, is showing rather frequent outbursts from B~20 to B=15.2 and is likely a dwarf nova of UGSS subtype. The other variable, USNO-B1.0 1321-0397655, has only one observed outburst in 1993 from R~19 to I=11.8 and is either a UGWZ dwarf nova or a recurrent nova. In both cases its next outburst can occur in the nearest future.

In course of the search for the new cataclysmic variables I have found the star USNO-B1.0 1320-0390658 with the coordinates $19^h21^m44^s.232$, $+42°04'41''.81$ (J2000.0) showing a strong variability between two Palomar epochs: B1=19.75, R1=19.14, B2=20.80, R2=15.87, I=18.61. Since this position falls at the crossing of three POSS-II areas, there are three $2^{nd}$ epoch plates in each band (Blue, Red and Infrared), as well as two $1^{st}$ epoch Red plates and two Quick-V photovisual plates. The star was just outside the field of view of the $1^{st}$ epoch Blue plate taken on 1951 Sep. 02.

The variable was at outburst on the 1991 Sep. 06 plate and in quiescence on all the remaining 13 plates. Fortunately, the $R$ magnitude for USNO-B1.0 catalogue was measured from the 1991 plate. The comparison of 1991 and 1995 Red plates centered at the new variable is shown in Fig. 1.

While examining the Palomar plates of this field I have serendipitously found another variable about 5′ North of the first one with a bright outburst on the 1993 June 11 Infrared plate. This star is also included into USNO-B catalogue as USNO-B1.0 1321-0397655 with the following coordinates and peculiar magnitudes: $19^h21^m48^s.934$, $+42°09'46''.60$, B1=20.48, R1=19.14, B2=20.61, R2=19.01, I=11.75. The comparison of 1993 June 11 and June 30 Infrared plates centered at the second variable is shown in Fig. 2.

The image of star in outburst was carefully examined to prove it is not an artifact. It has the same profile as that of other stars of similar brightness. The formal image characteristics (FWHM=8.7, flatness=0.03, maximum pixel value=23000) are matching other stars in 11-13$^m$ range. Even the profile asymmetry is the same (top of Gaussian is tilted to the West for all overexposed stars, including the variable). Finally, the coordinates of the star on June 11 plate are identical to those measured from all other images within the astrometric errors (0.1-0.2} for different plates). This perfect match makes it extremely unlikely to be a chance alignment of the star with a satellite glint or any other phenomenon, be it natural of artificial.

The 10′×10′ finder chart with the positions of both variables is presented in Fig. 3. The 1$^{st}$ variable is marked with two horizontal dashes, the 2$^{nd}$ - with a horizontal and a vertical dash. Both stars are not present in AAVSO Variable Star Index (VSX, Watson et al., 2006), the nearest variable listed there is NSVS 5539153 ($19^h21^m34^s.67$, $+42°03'03''.1$) - eclipsing variable of EW type with P=0$^d$.42079. The position of NSVS 5539153 is marked on the chart by an arrow. Also shown are Blue (USNO-A2.0) magnitudes of the comparison stars used in estimating new variables on Moscow archival plates.

Following the numbering scheme introduced back in 2007, the new variables were designated DDE 20 and DDE 21, using AAVSO observer code for the author. The list of variables discovered by DDE with their coordinates, finder charts and references is available online at http://hea.iki.rssi.ru/~denis/VarDDE.html.

The magnitudes of variables measured from Palomar plates are given in Table 1. It is notable that DDE 21 was fading after the outburst in June 1993. Its brightness on June 24 Red plate was ~1.5$^m$ above the quiescent level. Such fading is consistent with the outburst of the WZ Sge type dwarf novae which typically have outburst amplitudes about 7.5$^m$. However, the recurrent nova outburst cannot be excluded, as well.

**Table 1.** Magnitudes of new variables on Palomar plates.

| Date | Band | DDE 20 | DDE 21 |
|------|------|--------|--------|
| 1951 July 09 | Red | 19.14 | 19.14 |
| 1951 July 09 | Blue | 18.92 | 19.75 |
| 1951 Sep. 02 | Red | 18.97 | 18.49 |
| 1982 May 23 | V | <19.00 | 18.87 |
| 1982 May 23 | V | 19.10 | 18.72 |
| 1988 July 09 | Blue | 20.25 | 18.94 |
| 1989 July 04 | Blue | 20.80 | 19.29 |
| 1990 July 18 | Blue | 20.11 | 19.49 |
| 1991 Sep. 06 | Red | **15.87** | 18.92 |
| 1992 May 28 | IR | <18.50 | 18.20 |
| 1993 June 11 | IR | 18.61 | **11.75** |
| 1993 June 24 | Red | 19.32 | 17.54 |
| 1993 June 30 | IR | <18.50 | 17.95 |
| 1995 June 20 | Red | 19.27 | 18.36 |

To search for the possible past outbursts I have checked the existing photographic plates of Sternberg Astronomical Institute at MSU (Moscow plate collection, Samus et al., 2005). The new variables are inside 10°×10°FOV of 3 areas centered at 16 Lyr, 1933+39 and 1906+38.5 (η Lyr). The plates were obtained with 40-cm astrograph at Crimean Laboratory of SAI and have typical exposure times of 45-60 minutes with some exceptions towards the shorter side (10-30 min). Depending on sky conditions, exposure times and image quality, the limiting magnitudes of plates are varying in a wide range, being sometimes as good as 17.5, but typically 16.5-17.0. In total, 315 plates have been examined.

11 plates from Aug. 1957 centered at 1906+38.5 were not useful for estimating DDE 20 since they had 10-min exposures and 15$^m$ limit. On 4 plates the variables were outside the FOV, 2 plates had bad focus, 3 were taken through glass and 2 had trailed images. All the remaining plates were useful, even though the variables being located just 1.5 cm North and 3.5 cm West of the 16 Lyr field corner (plates are 30×30 cm$^2$).

In total, eight outbursts of DDE 20 were found on these plates with the maximum brightness $B$=15.2. Three outbursts have good coverage with 5-10 plates taken during several nights, while the other five were only detected on one plate each due to gaps in observations. The entire light curve of DDE 20 is shown in Fig. 4, the gray triangles being upper limits (rounded to 0.5$^m$). Since the magnitudes of the variable in outburst were estimated by visual inspection, typical errors are 0.1-0.2$^m$. Fig. 5 shows the well covered April 1968 outburst in more detail.

Dates of DDE 20 outbursts are listed in Table 2. The intervals between outbursts and their durations most likely correspond to the dwarf nova of SS Cyg subtype (UGSS), however the SU UMa (UGSU) classification is also possible.

**Table 2.** Outbursts of USNO-B1.0 1320-0390658 = DDE 20.

| Plate Number | Date dd.mm.yyyy | Time, UT | Exposure (min) | Mag |
|---|---|---|---|---|
| A05404 | 11.08.1967 | 17:54.0 | 45 | 16.1: |
| A05405 | 11.08.1967 | 18:43.0 | 45 | 16.0: |
| A05406 | 11.08.1967 | 19:29.0 | 45 | 16.1: |
| A05407 | 11.08.1967 | 20:15.0 | 45 | 15.9: |
| A05408 | 11.08.1967 | 21:01.0 | 45 | 16.2: |
| A05409 | 12.08.1967 | 18:24.0 | 45 | 15.2 |
| A05410 | 12.08.1967 | 19:10.0 | 45 | 15.3 |
| A05411 | 12.08.1967 | 19:56.0 | 45 | 15.4 |
| A05412 | 12.08.1967 | 20:42.0 | 45 | 15.6 |
| A05413 | 12.08.1967 | 21:28.0 | 45 | 15.5: |
| | | | | |
| A05751 | 18.04.1968 | 18:08.7 | 30 | 15.5: |
| A05755 | 18.04.1968 | 20:41.7 | 30 | 15.3 |
| A05766 | 19.04.1968 | 19:56.0 | 40 | 15.4 |
| A05767 | 19.04.1968 | 20:42.6 | 30 | 15.4 |
| A05781 | 21.04.1968 | 20:01.5 | 30 | 15.9 |
| A05782 | 21.04.1968 | 20:38.0 | 20 | 15.9 |
| A05795 | 22.04.1968 | 20:58.1 | 35 | 16.3 |
| A05796 | 22.04.1968 | 21:40.1 | 20 | 16.1 |
| A05805 | 23.04.1968 | 21:23.0 | 30 | 16.5: |
| A05806 | 23.04.1968 | 21:55.0 | 30 | 16.5: |
| | | | | |
| A06072 | 17.09.1968 | 17:05.0 | 47 | 15.8 |
| | | | | |
| A14464 | 22.08.1981 | 20:14.0 | 48 | 15.3 |
| | | | | |
| A15695 | 04.08.1983 | 21:16.1 | 60 | 15.5: |
| | | | | |
| A16310 | 03.05.1984 | 20:42.3 | 45 | 15.2 |
| A16311 | 03.05.1984 | 21:32.7 | 45 | 15.4 |
| A16312 | 03.05.1984 | 22:23.6 | 45 | 15.3 |
| A16313 | 03.05.1984 | 23:12.8 | 45 | 15.2 |
| | | | | |
| A18383 | 13.05.1988 | 22:09.4 | 60 | 15.2: |

No outburst of USNO-B1.0 1321-0397655 = DDE 21 was found on Moscow plates. Unfortunately there are no plates available between 1991 May 23 and 1995 July 27. The 2MASS infrared images were also checked. *JHK* images were taken on 1998 June 08, there is nothing at the position of the variable down to ~17.5 limiting magnitude. Together with USNO-B1.0 *B* and *R* magnitudes (*B-R* formally equals to 1.3-1.6) and faintness of the star on two POSS IR plates this excludes the possibility of DDE 21 being a red flaring star. As it was already mentioned above, this variable is a likely representative of WZ Sge dwarf novae type or even a yet more rare class of recurrent novae. Objects of both types are typically undergoing outbursts once in every 20-30 years. Thus, DDE 21 may be on the way to its next outburst at any time in the nearest future. The monitoring campaign of these two objects located within just 5' from each other is encouraged.

**Acknowledgements:** I would like to thank N. N. Samus, S. V. Antipin and E. V. Kazarovets for their long-time assistance in my work with Moscow plate collection and useful discussions.

**References:**

Samus, N. N., Sat, L. A., Zharova, A. V., 2005, in Proc. of Virtual Observatory Conf. on Plate Content Digitization, April 27-30, 2005, Sofia, Bulgaria

Watson, C., Henden, A., Price, A., 2006, in Proc. of Soc. Astro. Sci. Symposium on Telescope Science, May 23-25, 2006, Big Bear, CA

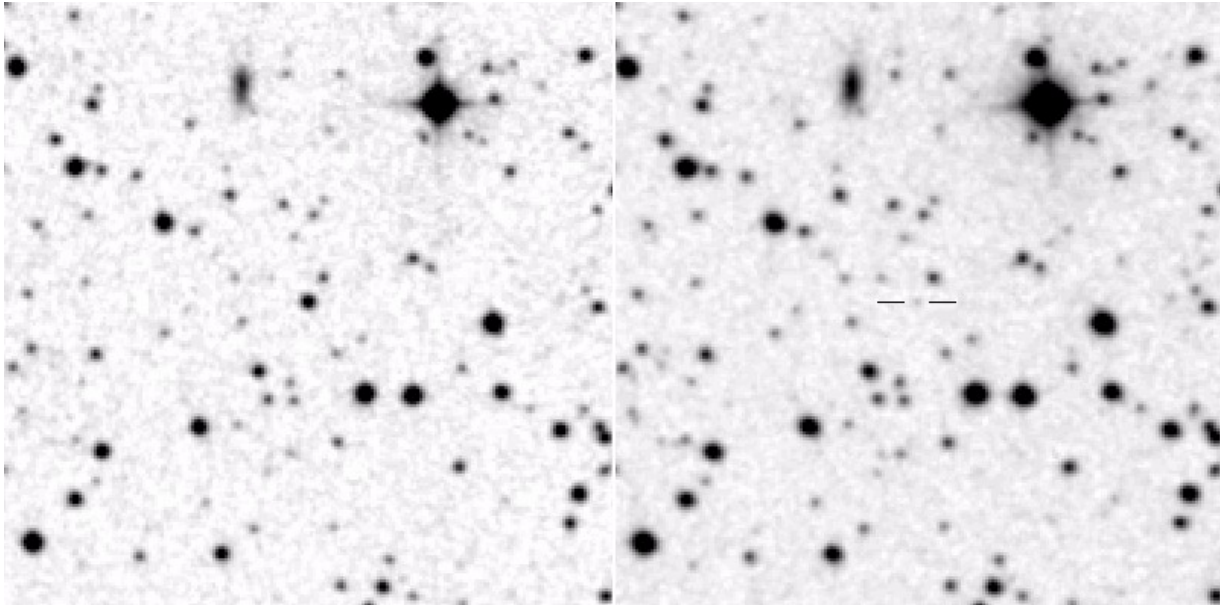

**Figure 1.** USNO-B1.0 1320-0390658 (DDE 20) on the Red Palomar plates taken on 1991 Sep. 06 (left) and on 1995 June 20 (right). FOV is 200″×200″, North is up, East is to the left.

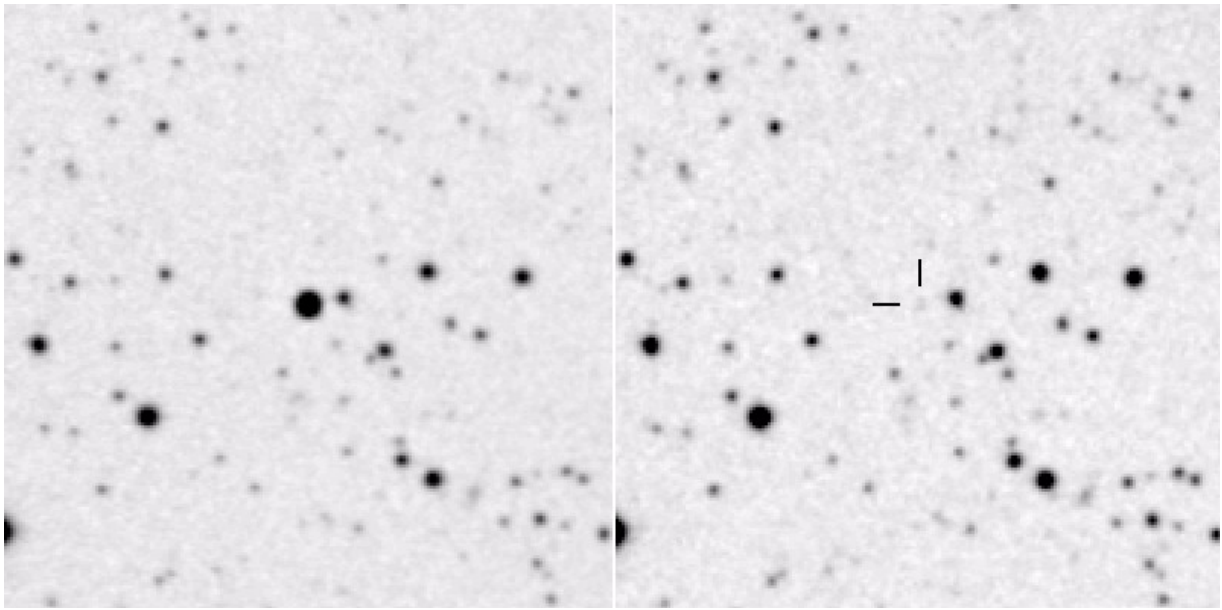

**Figure 2.** USNO-B1.0 1321-0397655 (DDE 21) on the Infrared Palomar plates taken on 1993 June 11 (left) and on 1993 June 30 (right). FOV is 200″×200″, North up, East left.

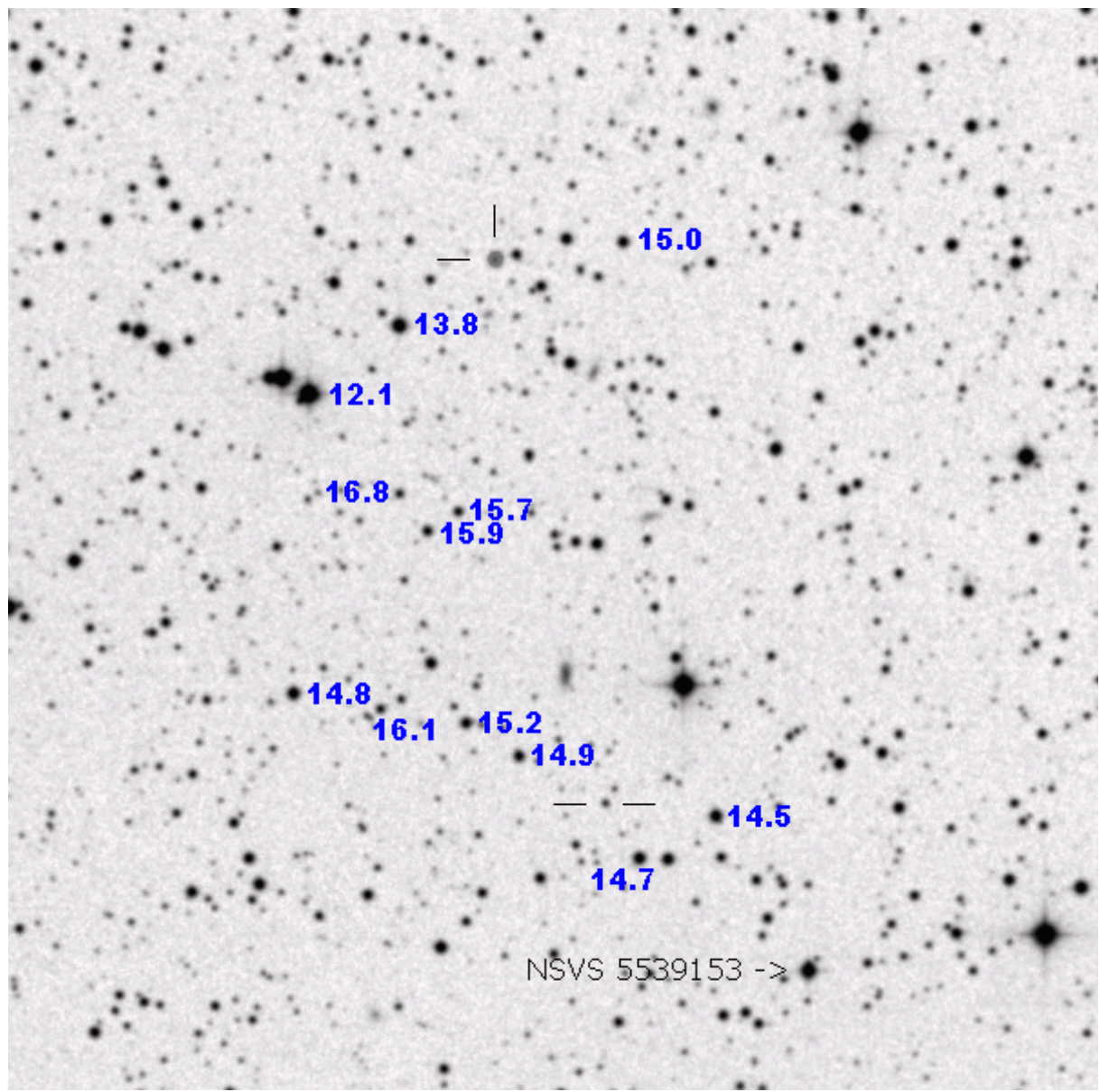

**Figure 3.** 10'×10' finder chart of two new variables. North up, East left.

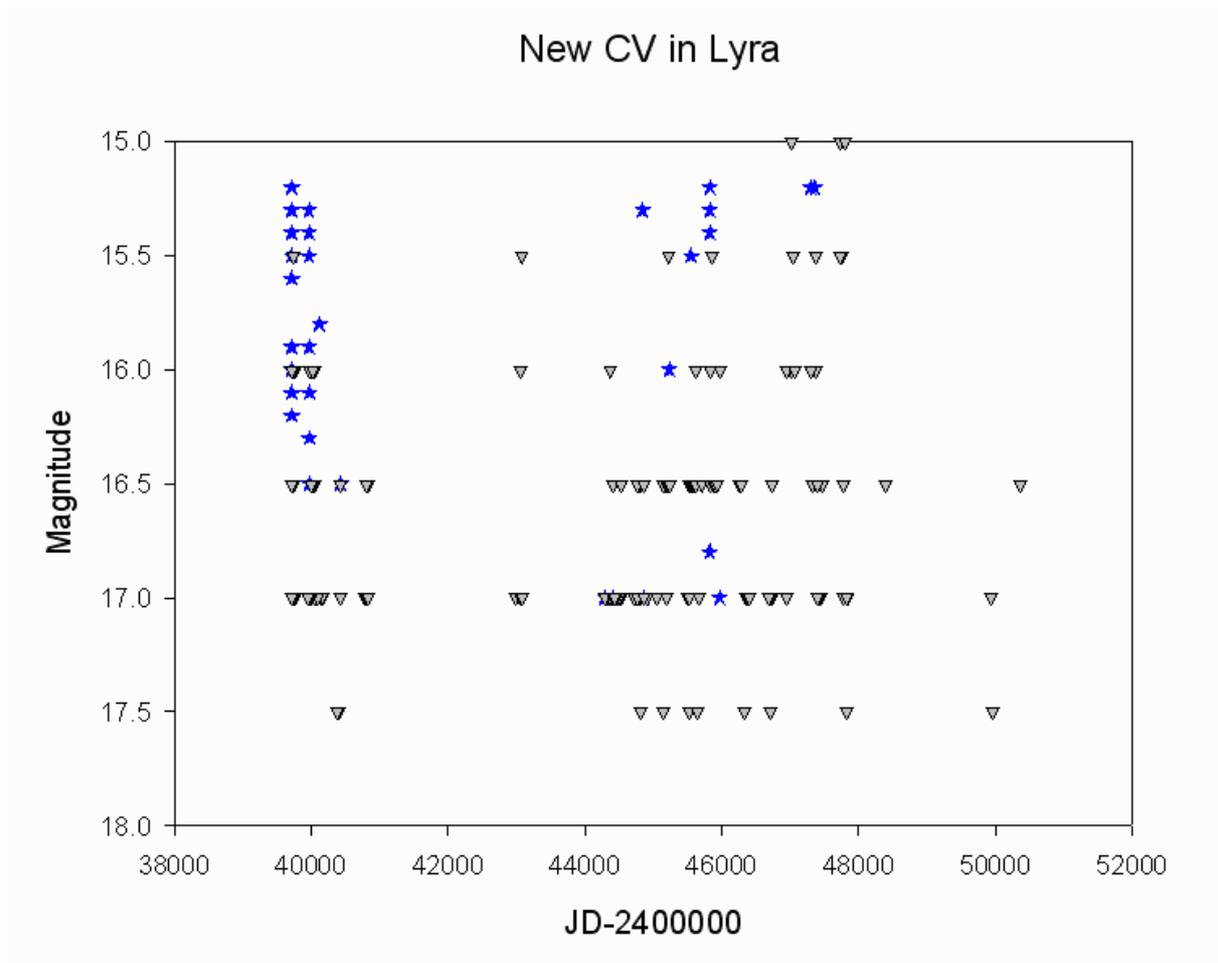

**Figure 4.** Light curve of DDE 20 from Moscow photographic plates. Blue stars - positive detections, gray triangles - upper limits.

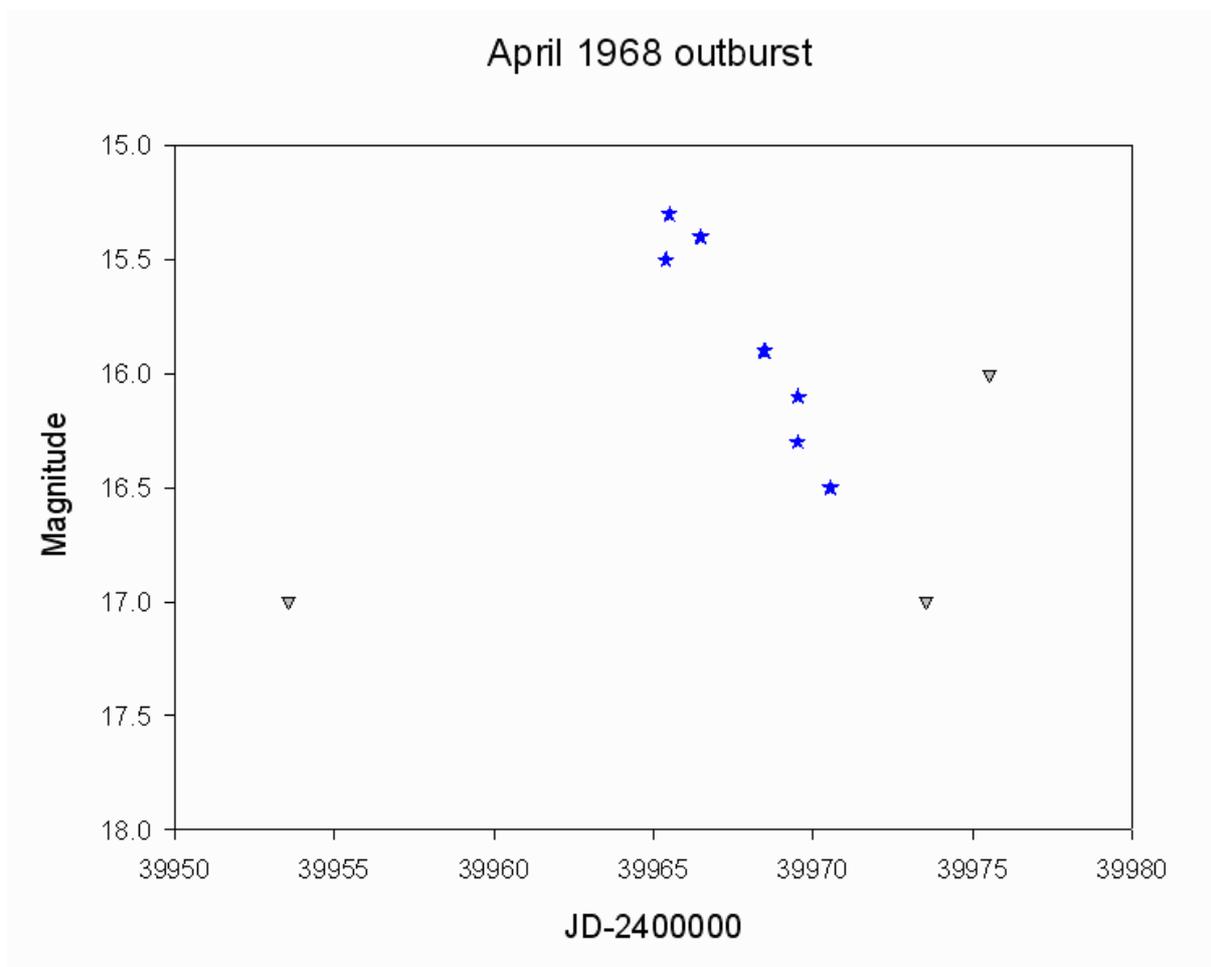

**Figure 5.** Light curve of April 1968 outburst of DDE 20 from Moscow photographic plates.